# Engineering Collaborative Social Science Toolkits

## STS Methods and Concepts as Devices for Interdisciplinary Diplomacy

Peter Müller and Jan-Hendrik Passoth

**Abstract** The smartification of industries is marked by the development of cyber-physical systems, interfaces, and intelligent software featuring knowledge models, empirical real-time data, and feedback-loops. This brings up new requirements and challenges for HMI design and industrial labor. Social sciences can contribute to such engineering projects with their perspectives, concepts and knowledge. Hence, we claim that, in addition to following their own intellectual curiosities, the social sciences can and should contribute to such projects in terms of an 'applied' science, helping to foster interdisciplinary collaboration and providing toolkits and devices for what we call 'interdisciplinary diplomacy'. We illustrate the benefits of such an approach, support them with selected examples of our involvement in such an engineering project and propose using methods as diplomatic devices and concepts as social theory plug-ins. The article ends with an outlook and reflection on the remaining issue of whether and in how far such 'applied' and critical social science can or should be integrated.

## 1 Social Science in Engineering Projects

The transition from traditional to smart industries is indicated by the implementation of (digital) automation tools and the integration of cyber-physical systems, e.g. sensory interfaces, informed models and (self-learning) algorithms. This refurbishment of factory (infra)structure is affecting HMI design decisions and the ways that such smartified plants are operated [Pos15]. Industrial smartification thus involves dense entanglements of business, organization, technology and labor-routine issues, all of which are of genuine interest to disciplines like sociology and "Science & Technology Studies" (STS). In the context of a deep involvement in an engineering project on smart factories and industrial automation, we as STS researchers have composed a qualitative methods toolbox to help understand the effects of a transition from traditional to smart production. In section 3 of this paper, we will give a very short overview of this toolbox and how we used it in the project. We did this not only for the sake of a sociological understanding (of socio-technical arrangements) but also for engineering purposes of human machine interface (HMI) design as well as to acquire technological, formal knowledge models. Similar methodologies have already been designed and elaborated in specific engineering disciplines, especially by those who work in the field of HMI [Jan16]. In HMI, but also other engineering fields, methods from cognitive science and ergonomics and also from the social sciences (in a way, since they are socio-scientific methods being used for different purposes) are commonly employed in order to acquire (expert) knowledge, develop mental models and assess suitable and supportive interface designs [Liu10]. By providing a toolbox, we further contribute by adopting sociological methods not limited to cases in which social science project members demand such a methodology. We will also illustrate how such a methodology might contribute to the methodological repertoire of this specific field of engineering, turning our methods into devices for interdisciplinary diplomacy.

To illustrate the usefulness of such an approach, we used this toolbox and a set of STS concepts to develop recommendations for the training of industrial plant operators using several generic training scenarios designed specifically for smart factory HMI cases. We will highlight some of these recommendations in section 4 of this paper. Our goal was to come up with a broader framework to constantly revise and update the training scenarios that have already been applied and also to train supervising staff in how to use the data generated thus far to sort out common patterns of successful plant operation that go beyond intrinsic plant models and practical operating guidelines like component maintenance. The purpose of such an approach is to help guide the reconfigurations of socio-technical arrangements in complex work environments that are at the core of transitions from traditional to smart production. Such a transition, we conclude, must be careful – following Annemarie Mol's notion of care [Mol08] in the sense that it needs care and must be taken care of – and we must be incremental to ensure it is done in a responsive and responsible way that helps to increase both the autonomy and the self-determination of operators and supervising staff.

We will conclude this piece with a short reflection (section 5) on the affirmative or critical nature of such an approach to the integration of social science research in engineering projects. We will argue that while it is true that such deep integration in projects makes it tricky to criticize their overall ends, it is only through such immersion that it is possible to produce concrete alternatives and challenge taken-for-granted assumptions in engineering.

## 2 Smart Factories and Smart Collaborations

As partners of the EU-funded project 'IMPROVE', we investigated the 'socio-technical aspects' of the project. IMPROVE is focused on developing automation technology for smart factories, improving plant surveillance for operators, and providing self-learning software that contains plant models and processes real-time data which are used for early anomaly detection and malfunction anticipation. But how do we as social scientists and STS researchers function as a part of this project? We worked very close with the HMI engineers from IMPROVE who were not only commissioned with designing an interface for surveillance and detection features, but also with developing a decision support system (DSS) for plant operators in this and other smart factory contexts. Furthermore, their work package involved the task of eliciting expert informed mental models of the industrial project partners' plants. To do this, they used card-sorting techniques and interview sessions with the technical and operating personnel from IMPROVE's industrial partners. From this they were able to build ordinal cause-effect graphs to be used for operating purposes in terms of monitoring and decision support. For our deliverable as social science partners, it was necessary to collaborate with the HMI project team in order to understand their design assignments and means and thus to contribute by providing socio-scientific consulting and complementary content like data or concepts. Our engineering partners shared their data (and resulting models and prototypes) and took us with them to their inquiry meetings. Based on these data and experiences we attempted to understand the capacities and underlying strategies of their methodology – an eclectic assemblage of psychological, ergonomic and social science methods. Furthermore, we could gather our own field data – on the investigated operators and our researcher colleagues and their very own social practices of engineering [Buc94] – and experimented with ways of analysis that could complement our partners' data analysis. While they used the data to create the plants' models, we have tried to figure out some characteristics of operators' labor routines that would be relevant to HMI design. As far as they could be reconstructed based on our dataset, we drafted concepts of how operators and supervisors handle problems, how they conceptualize their own practice, and how they configure their organizational roles.

These very socio-technical arrangements that we have mapped informed our additional tasks involving the development of a concrete, complementary toolkit of qualitative research methods that would fit such engineering projects. Hence, we not only acted as methodological consultants but, in particular, as methodological researchers who were tinkering with a particular toolbox that could inform HMI design and DSS features. Also, we experimented with generic training scenarios for operators in smart factories in ways that featured IMPROVE's or similar technology. We grounded both assignments in the same data and concepts in order to design an analytical setup for our toolbox and in order to develop training scenarios that took into account the operators' particular practices and the specific knowledges and skills that are required. Therefore, we have worked on a conceptualization for practice blueprints concerning the different types of problems operators encounter in their work. Our idea was that the HMI design and practices covered by the training schemes provided should be more integrated and could even feature reciprocal synergies.

We have also observed another differentiation of methodological research approaches that is seen less frequently in sociological studies. In addition to the qualitative methods applied by IMPROVE's HMI partners, other partners used data mining to receive quantified, explicit data and models that would feature precise predictions. The applied quantitative and qualitative methods of our engineering partners were decidedly focused on explicit correlations and configurations of data and models that were both quantitative and qualitative. Taking our own contribution into account, this adds up to a threefold (at least) methodological setup, where we provided research methods that mostly covered implicit, tacit knowledge and latent, subtle orders of practice. While our partner oriented their work towards use-cases, we were rather looking for 'problem-cases', e.g. unresolved tensions of interest, places of interference between engineering and worker (operator) mindsets. We prepared a methodological toolbox because we wanted not only to contribute additional data as social science experts but also to add a perspective that could cross and integrate our partners' quantitative and qualitative approaches and findings. To do so, it was necessary to re-think our roles as methodologists: (how) could we (re)invent ourselves as socio-engineers of diplomatic devices within interdisciplinary engineering projects?

## 3 Methods as Diplomatic Devices

Qualitative research methods are already known within the fields of engineering, especially digitalization. However, from a sociological point of view, these methods are not sufficiently elaborated – or at least not properly implemented and justified – by their users. In many publications that deal with such methodological needs or issues, corresponding methods are applied but not explained. In Software Engineering, this has been recognized and initial steps have been explicitly taken to turn towards more sound empirical research [Dit08], [Tor11], [Han07], but such an approach is missing from industrial engineering. Nonetheless, this must not be mistaken for some kind of sociological snobbism. This is not about pushing sociological questions into other disciplines, but clearing up what such qualitative methods are capable of, and how their application can even be used to help produce an understanding of disciplinary boundaries and ways of crossing them. As a toolbox for collaborative projects, they can be thought of as *diplomatic devices*. We have turned several social science methods into such devices. In particular, we have focused on the qualitative analysis of technical documents and on ethnographic fieldwork. A textual version of this toolbox has been created for our project deliverable and will soon be published as part of collaboratively edited collection [Pas18].

In engineering contexts, these methods – although in most cases it is only interviews – are used as knowledge-acquiring methods that are focused on objectives. These methods are, or at least seem to be, rather formalized, explicit and objective (insofar as they describe objects' qualities). As a result, such methods mostly consist of cognition science methods, ergonomics, and (complementary) interviews [Han07], [Jan16]. These interviews, in particular, seem to back up the other methods of model generating and literary review. Interviews, however, are not a standard procedure only within the humanities and social sciences. The sociological style of doing interviews, however, is more elaborate in certain respects. This concerns several methodological principles and guidelines, e.g. CA transcription (conversation analysis transcript or CAT) [Sch73] or the several interview guideline revisions (whether it is open, closed or structured) due to their pretesting results. From an engineering perspective, these methodological norms might seem very exaggerated. Furthermore, since engineering assignments were met with less methodological effort when it comes to using interviews for requirements engineering and testing purposes, one might be tempted to agree with this assessment. However, we will give several arguments to the effect that these efforts do, in fact, pay off scientifically and in terms of technology development and design. Integrating social science researchers and social science methodology enables a project to ground its work in a common understanding whose quality satisfies all its epistemological and technical requirements and which is common because it has been established collaboratively. For example, it is possible to ground the engineering of formal models, HMI and socio-scientific reflections on the same empirical data, thus increasing the capacity for (and likelihood of) synergetic exchange between different researchers and of the overall coherence within such a project.

In the course of the project, we were able to accompany our engineering colleagues on requirements engineering visits, provide methodological consultation and host data analysis sessions. We attempted to turn the methodological canon and controversies of social science interview research into a toolbox equipped with a heterogeneous set of devices, we assumed that the organizing principle of such a collection of devices does not need to be rigor and coherence, but rather its usefulness. It was designed to be useful for the common project of treating the various actors we encountered (operators, managers, industrial researchers and, yes, ourselves) as part of a 'public' as John Dewey understood it [Dew06]: as something that "cannot be mastered by anyone but that can be represented, over and over again, by the social sciences and the humanities" [Lat03]. This might also be identified as a Meadian institution, for it addresses "situations which we admit are not realized but which demand realization" [Mea23], thus are to be handled constantly in an infinite struggle of methodological feedback. This toolbox – a collection of revisable how-tos, visits and workshops – was the basis for our own substantial contribution to the overall project (see section 5 for a short reflection on this). It was also the basis for unpacking a controversy dealing with claims of validity, epistemic authority and pragmatic usefulness instead of just glossing over the differences between disciplinary cultures by proposing a methods-based consensus.

## 4 Social Theory Plug-Ins

As partners in the project consortium, our task as STS researchers was not only to encourage interdisciplinary research and help our engineering partners talk to actors they encounter in the empirical context of industrial research and practice, but also to provide recommendations on how to deal with the reconfiguration of sociotechnical arrangements that are at the practical core of the transition from traditional to smart production. Our recommendations for dealing with the reconfiguration of sociotechnical arrangements are prototypes because the work of our engineering partners, for example, on the use of machine learning for semi-automatic alarm analysis or on the prototype for a self-adapting HMI, are also proof-of-concept demonstrators and prototypes rather than concrete implementations of a new version of a marketable production facility. We explicitly used three *social theory plug-ins*: a relational and procedural concept of agency based on pragmatism [Dew96], [Mea03], social science approaches to implicit knowledge based on practice theory [Col01] and a conceptual framework for human-machine cooperation based on Actor-Network Theory [Pas15]. They provided the ground for our design and training recommendations, we used them to challenge and rework the mostly implicit, but sometimes also very explicit, assumptions about work and automation, tacit knowledge, and HMI principles used for modeling and design by our engineering partners. As in the case of the toolbox described above, these plug-ins served a double purpose for us: they are at the same time provisional results of STS research in an engineering project as well as ways of intervening in the daily work of engineering practice. In this way, they open up already closed (and sometimes too quickly closed) debates about goals and work packages and propose alternatives and collaboratively develop ways of dealing with the changes in sociotechnical arrangements introduced by the project as a whole [Jen01]. They are "lateral concepts" that enable "ontological experiments" [Gad16], [Jen15].

One example of this approach is the following: By using these *social theory plug-ins*, we argued, in design meetings and in comments on requirements and models, that HMIs should not be regarded as one-way tools for monitoring and controlling machines because operators will then be required to solve the (nearly impossible) problems of translating their implicit knowledge into explicit machine instructions and of mapping (standardized) HMI features onto (tacitly) known routines and patterns of trained behaviour. On the contrary, we suggested that operators should at least be able to organize themselves through the HMI and reflect upon the HMI's role in their practices in case of suboptimal operating processes. This is no mere rejection of operator responsibility, which, since it plays a significant role in daily work should still remain on the operator's side. Rather, it enables a different way of accounting for best practices, work-arounds, glitches and failures in any current and future HMI design. However, operators do not just use HMI. Training can focus on supporting operators to enhance themselves through the HMI, which will also help to resolve responsibility and decision dilemmas in case an integrated decision support system contradicts the operator's intuition. We therefore suggested that operator trainings need to focus on three issues: a symmetrical approach to HMI that enhances human-machine cooperation and adaptation in both directions, classifying incidences by whether they require either implicit or explicit knowledge and managing their (and other operators') knowledge and experience through the HMI by giving feedback on provided information and recommended interventions as well as by reporting and storing their own knowledge and experiences. Social theory of knowledge in practice and actor-networks can thus be used to organize information, classify incidences and provide an analytic framework for further, recursive adaptations [Ber98].

## 5 Involvement and Intervention

The interdisciplinary approach we have presented here goes beyond the mere contribution of extra-technological contexts like marketing, organization, policy, technology assessment or 'nice to have' contemplations. To reflect on the implicit premises of technological developments or on the societal meaning of their implementation is usually regarded as a genuine and specifically sociological duty. While, from an engineering point of view, such contemplation might rather appear as an ornament of interdisciplinary projects, it is exactly what social scientists regard as their primary obligation, thus those often harshly criticize colleagues who engage with instrumental, affirmative tasks in such interdisciplinary projects for doing so.

But are 'applied' and reflexive social science approaches, after all, mutually exclusive options [Hor02]? Collaborations between social scientists and engineers require the social scientist, indeed, to get her- or himself into the technical, instrumental setup of the engineering project – and that means, at least for the sake of cooperation, accepting the project's frame of reference and affirming the project's cultural, economic and organizational premises, established facts and assumptions. As a result, social science critiques are constrained by the explicit ends defined by the project, and radical critique and reflections on the *conditions of possibility* of such social situations are rendered impossible. Although social scientists do not need to completely narrow their perspectives, concerns and issues regarding the project's framework, collaboration might yet compromise their capacities for social criticism. This is a concern that causes many social scientists to reject 'applied' science scenarios of social science. It is, of course, quite possible for social scientists to add subaltern interests and critiques to the tenor of what is taken into account concerning technological design. For example by reflecting on diversity and thus inspiring more inclusive interfaces or devices (e.g. airbags positioned to consider physiognomic gender disparities) or helping to design software that incorporates organizational responsibility and accountability distributions or that helps to eliminate hierarchical tensions (e.g. bi-directional communication or feedback loops instead of mere monitoring and intervening). In this way, immediate problems are resolved, but the structural sources of these problems remain intact and might even (re)appear in a hardened form. After all, collaboration can also be a pejorative term, and it thus holds on to this residual connotation despite its recent popularity in terms of interdisciplinarity [Nie14].

To conclude, is the integration of applied and critical social science designed to fail? It is unquestionable that, however such integration is done, it can neither avoid trade-offs (on the instrumental or critical side) nor replace proper, exclusive social criticism. Nevertheless, interdisciplinary collaborations between social scientists and engineers have two advantages: on the one hand, they offer a deep insight into engineering cultures, in terms of engineers' working culture and the cultural significance of technological artifacts with respect to their demands and effects. On the other hand, such collaborations enable social scientists to provide interdisciplinary diplomacy, to mediate and offer consultation within their project and with regard to its social context. Without forgetting about the aforementioned critique conundrum, both features meet certain aspirations of social criticism: to get involved and be in touch with social situations, to contribute tangible and actual (critical) interventions. This avoids the separation of practical and intellectual work corresponding to social segregation and stratification [Hor02]. Eventually, if "technology is society made durable" [Lat90], the usual critical practice of watching and judging from afar is more than unacceptable because the question of which society is made durable and how is a never-ending, substantial concern. Interdisciplinary diplomacy is a way to share and spread this concern within interdisciplinary collaborations and to start working (together) on concrete alternatives.